# Multifractal Fluctuations in Seismic Interspike Series

Luciano Telesca* and Vincenzo Lapenna

Istituto di Metodologie per l'Analisi Ambientale, Consiglio Nazionale delle Ricerche, C.da S.Loja, 85050 Tito (PZ), Italy

Maria Macchiato

Dipartimento di Scienze Fisiche, INFM, Università Federico II, Naples, Italy

**Abstract**

Multifractal fluctuations in the time dynamics of seismicity data have been analyzed. We investigated the interspike intervals (times between successive earthquakes) of one of the most seismically active areas of central Italy by using the Multifractal Detrended Fluctuation Analysis (MF-DFA). Analyzing the time evolution of the multifractality degree of the series, a loss of multifractality during the aftershocks is revealed. This study aims to suggest another approach to investigate the complex dynamics of earthquakes.

Corresponding author: ltelesca@imaa.cnr.it

## 1. INTRODUCTION

Earthquakes belong to the class of spatio-temporal point processes, marked by the magnitude. Power-law statistics characterize their parameters. The Gutenberg-Richter law states that the probability distribution of the released energy is a power-law [1]. The epicentres occur on a fractal-like distribution of faults [2]. The Omori's law states that the number of aftershocks, which follow a main event, decays as a power-law with exponent close to minus one [3]. The fractal behavior revealed in these statistics could be considered as the end-product of a self-organized critical state of the Earth's crust, analogous to the state of a sandpile, which evolves naturally to a critical repose angle in response to the steady supply of new grains at the summit [4]. In recent studies, it has evidenced that characterizing the temporal distribution of a seismic sequence is an important challenge. The interspike intervals (time between two successive seismic events) follow a Poissonian distribution for completely random seismic sequences, while they are generally power-law distributed for time-clusterized sequences [5]. Time-clusterized seismic sequences are featured by time-correlation properties among the events, contrarily to Poissonian sequences, which are memoryless processes. But the probability density function (pdf) of the interspike intervals is only one window into a point process, because it yields only first-order information and it reveals none about the correlation properties [6]. Therefore, time-fractal second-order methods are necessary to investigate the temporal fluctuations of seismic sequences more deeply. The use of statistics like the Allan Factor [7], the Fano Factor [8], the Detrended Fluctuation Analysis (DFA)[9], has allowed getting more insight into the time dynamics of seismicity [10]. All these measures are consistent with each other, so that we can define one scaling exponent that is sufficient to capture the time dynamics of a seismic process.

But one scaling exponent is sufficient to completely describe a seismic process under the hypothesis that this is monofractal. Monofractals are homogeneous objects, in the sense that they have the same scaling properties, characterized by a single singularity exponent [11]. The need for more than one scaling exponent can derive from the existence of a crossover timescale, which separates regimes with different scaling behaviors [12], suggesting e.g. different types of correlations at small and large timescales [13]. Different values of the same scaling exponent could be required for different segments of the same sequence, indicating a time variation of the scaling behaviour, relying to a time variation of the underlying dynamics [14]. Furthermore, different scaling exponents can be revealed for many interwoven fractal subsets of the sequence [15]; in this case the process is not a monofractal but multifractal. A multifractal object requires many indices to characterize its scaling properties. Multifractals can be decomposed into many-possibly infinitely many-sub-sets characterized by different scaling exponents. Thus multifractals are intrinsically more complex and inhomogeneous than monofractals [16] and characterize systems featured by a spiky dynamics, with sudden and intense bursts of high frequency fluctuations [17].

A seismic process can be considered as characterized by a fluctuating behaviour, with temporal phases of low activity interspersed between those where the density of the events is relatively large. This “sparseness” can be well described by means of the concept of multifractal.

The simplest type of multifractal analysis is given by the standard partition function multifractal formalism, developed to characterize multifractality in stationary measures [18]. This method does not correctly estimate the multifractal behaviour of series affected by trends or nonstationarities. Another multifractal method was the so-called wavelet transform modulus maxima (WTMM) method, based on the wavelet analysis and able to characterize the multifractality of nonstationary series. This method involves tracing the maxima lines in the continuous wavelet transform over all scales, and from the point of view of estimating the "local" multifractal structure in time series, the WTMM method permits to quantify the local Hurst exponent and its variation in time. In fact, the WTMM methods allows quantifying and visualizing the temporal structure of the different fractal sub-sets, characterized by different local Hurst exponents and different density [19]. The only disadvantage of WTMM method is a big effort in programming.

Alternatively, a method based on the generalization of the detrended fluctuation analysis (DFA) has been developed [20]. This method, called multifractal detrended fluctuation analysis (MF-DFA) is able to reliably determine the multifractal scaling behavior of nonstationary series, without requiring a big effort in programming.

## 2. OBSERVATIONAL SEISMICITY DATA

We study the earthquake sequence of a very seismically active area of central Italy, which was struck by a violent earthquake ($M_D$=5.8) on September 26, 1997. The epicentre distribution of the events occurred from 1983 to 2003 is shown in Fig. 1 (data extracted from the instrumental catalogue of National Institute of Geophysics and Volcanology-INGV [21]). The earthquakes are located in a circular area centered on the epicenter of the strongest M5.8 earthquake, with a radius of 100 km. The area is consistent with the recent seismotectonic zoning of Italy [22]. The completeness magnitude, estimated after performing the Gutenberg-Richter analysis [1], is 2.4. Fig. 2 shows the interspike interval series.

## 3. THE MULTIFRACTAL METHOD

The main features of multifractals is to be characterized by high variability on a wide range of temporal or spatial scales, associated to intermittent fluctuations and long-range power-law correlations.

The interspike intervals examined in this paper present clear irregular dynamics (Fig. 2), characterized by sudden bursts of high frequency fluctuations, which suggest to perform a multifractal analysis, thus evidencing the presence of different scaling behaviours for different intensities of fluctuations. We applied the Multifractal Detrended Fluctuation Analysis (MF-DFA), which operates on the series x(i), where i=1,2,...,N and N is the length of the series. With $x_{ave}$ we indicate the mean value

$$x_{\text{ave}} = \frac{1}{N}\sum_{k=1}^{N} x(k) \, . \quad (1)$$

We assume that x(i) are increments of a random walk process around the average $x_{ave}$, thus the “trajectory” or “profile” is given by the integration of the signal

$$y(i) = \sum_{k=1}^{i}\left[x(k) - x_{\text{ave}}\right]. \quad (2)$$

Next, the integrated series is divided into $N_S$=int(N/s) nonoverlapping segments of equal length *s*. Since the length N of the series is often not a multiple of the considered time scale s, a short part at the end of the profile y(i) may remain. In order not to disregard this part of the series, the same procedure is repeated starting from the opposite end. Thereby, $2N_S$ segments are obtained altogether. Then we calculate the local trend for each of the $2N_S$ segments by a least square fit of the series. Then we determine the variance

$$F^2(s,\nu) = \frac{1}{s}\sum_{i=1}^{s}\{y[(\nu-1)s+i] - y_\nu(i)\}^2 \quad (3)$$

for each segment ν, ν=1,..,$N_S$ and

$$F^2(s,\nu) = \frac{1}{s}\sum_{i=1}^{s}\{y[N-(\nu-N_S)s+i] - y_\nu(i)\}^2 \quad (4)$$

for ν=$N_S$+1,..,$2N_S$. Here, $y_\nu(i)$ is the fitting line in segment ν. Then, we average over all segments to obtain the q-th order fluctuation function

$$F_q(s) = \left\{\frac{1}{2N_S}\sum_{\nu=1}^{2N_S}\left[F^2(s,\nu)\right]^{\frac{q}{2}}\right\}^{\frac{1}{q}} \quad (5)$$

where, in general, the index variable q can take any real value except zero. Repeating the procedure described above, for several time scales *s*, $F_q(s)$ will increase with increasing s. Then analyzing log-log plots $F_q(s)$ versus s for each value of q, we determine the scaling behavior of the fluctuation functions. If the series $x_i$ is long-range power-law correlated, $F_q(s)$ increases for large values of s as a power-law

$$F_q(s) \approx s^{h_q}. \quad (6)$$

The value $h_0$ corresponds to the limit $h_q$ for $q \to 0$, and cannot be determined directly using the averaging procedure of Eq. 5 because of the diverging exponent. Instead, a logarithmic averaging procedure has to be employed,

$$F_0(s) \equiv \exp\left\{ \frac{1}{4N_S} \sum_{\nu=1}^{2N_S} \ln\left[F^2(s,\nu)\right] \right\} \approx s^{h(0}. \quad (7)$$

In general the exponent $h_q$ will depend on q. For stationary series, $h_2$ is the well-defined Hurst exponent H [23]. Thus, we call $h_q$ the generalized Hurst exponent. $h_q$ independent of q characterizes monofractal series. The different scaling of small and large fluctuations will yield a significant dependence of $h_q$ on q. For positive q, the segments ν with large variance (i.e. large deviation from the corresponding fit) will dominate the average $F_q(s)$. Therefore, if q is positive, $h_q$ describes the scaling behavior of the segments with large fluctuations; and generally, large fluctuations are characterized by a smaller scaling exponent $h_q$ for multifractal time series. For negative q, the segments ν with small variance will dominate the average $F_q(s)$. Thus, for negative q values, the scaling exponent $h_q$ describes the scaling behavior of segments with small fluctuations, usually characterized by larger scaling exponents. Since the MF-DFA can only estimate positive $h_q$, it becomes inaccurate for strongly anti-correlated series, which are characterized by $h_q$ close to zero. In such cases, the single summation of Eq. 2 has to be substituted by the double summation

$$\hat{Y}(i) = \sum_{k=1}^{i} \left[ y(k) - y_{\text{ave}} \right] \quad (8)$$

and the exponent is $h'_q = h_q + 1$, which accurately estimate $h_q$ less than zero, but larger than –1.

## 4. RESULTS

We performed the MF-DFA over the seismic interspike time series of central Italy (Fig. 1). We calculated the fluctuation functions $F_q(s)$ for scales s ranging from 10 events to N/4, where N is the total length of the series. The length of the series (N=7520) allows us to consider the estimated exponents reliable.

Fig. 3 shows the fluctuation functions $F_q(s)$ for q=-10 and q=+10. The different slopes $h_q$ of the fluctuation curves indicate that small and large interspike fluctuations scale differently. We calculated the fluctuation functions $F_q(s)$ for $q \in [-10, 10]$, with 0.5 step. Fig. 4 shows the q-dependence of the generalized Hurst exponent $h_q$ determined by fits in the regime $10 < s < N/4$. The $h_q \sim q$ relation is characterized by the typical multifractal form, monotonically decreasing with the increase of q. A measure of the degree of multifractality can be given by the $h_q$-range (maximum-minimum) $\rho_q$ or by the standard deviation $\sigma_q$, which are ~0.52 and ~0.20 respectively in the present case.

We investigated the time variation of the multifractality of the interspike series analysing the time variation of the set of the $h_q$ exponents. We calculated the set of the generalized Hurst exponent $\{h_q(t): -10 \leq q \leq +10\}$ in a $10^3$ event long window sliding through the entire series by one event. Each set is associated with the time of the last event in the window. The window length allows obtaining reliable values of the exponents; the shift between successive windows permits sufficient smoothing among the values. Fig. 5 shows the time variation of the generalized Hurst exponents $h_q(t)$, clearly characterized by a strong variability, and this suggests a strong variability of the multifractal character of the series. The projection on the plane $h_q$-t is shown in Fig. 6, where the upper and the lower curves are the envelopes of the highest (q=-10) and the lowest (q=10) Hurst exponents respectively. The variability range of the exponents is approximately constant up to the occurrence of the largest earthquake of the series (M=5.8), lowering during the aftershock activation. Fig. 7 shows a grey-scale map of the variability of the Hurst exponents $h_q$ as a function of the event number and the parameter q. After the main event (indicated by the red vertical line) the values of $h_q$ reduce their variability during the occurrence of the aftershocks. Fig. 8 shows the time variation of the range $\rho_q$ (Fig. 8a) and the standard deviation $\sigma_q$ (Fig. 8b) of $h_q$, which are approximately constant up to the occurrence of the mainshock and tend to zero after it; during the aftershocks both the parameters assume low values, while slightly increase after the phase of aftershock activation has finished. In order to check the loss of multifractality after the main event, we performed our analysis on surrogate time series. For each window, we generated ten surrogate interspike series by shuffling the original series. The new series preserve the distribution of the interspike interval distribution but destroy all the correlations among them. Hence the generated series are simple random walks, characterized by monofractal behavior. In Fig. 8 we also plotted the mean value($\pm 1\sigma$) over ten surrogate series of the range $\langle\rho_{q,shuffled}\rangle$ (Fig. 8a) and the standard deviation $\langle\sigma_{q,shuffled}\rangle$ (Fig. 8b); both curves assume values close to zero, as expected for monofractal series. This result validates that before and after the mainshock a significant change in the multifractal characteristics of the seismic series occurred.

## 5. CONCLUSIONS

Recently, the method of DFA [27] has shown its potential in revealing monofractal scaling behavior and detecting long-range correlations in noisy and nonstationary time series [11, 28] in many different scientific fields (genetics [27, 29], cardiology [30], geology [31], economics [32]).

The potential of multifractal analysis is far from being fully exploited, since it was only very recently that attention has been drawn to the need for a thorough testing of the multifractal tools, and, in particular, a much deeper understanding of the nature of the seismic phenomena and their interactions with the multifractal methods used for their analysis [14, 33].

The geophysical phenomenon underlying earthquakes is complex. The multifractal analysis, performed in the present study, has led to a better understanding of such complexity. The multifractality of a seismic phenomenon relies on the different scaling for long and short interspike intervals. The analysis of the time evolution of the multifractality degree, measured by the range or the standard deviation of the generalized Hurst exponents, suggests that the seismic system is characterized by a dynamical change from heterogeneity toward homogeneity during the aftershock activation, revealed by a loss of multifractality after a main event. In particular, the mechanisms of generating aftershocks can be of two different types [34]: 1) the mainshock does not release the stress completely, and some patches of the fault remain unruptured or the amount of slip is less than on the patch where the main event occurs; or 2) the main event modifies the stress field in the volume surrounding the fault [35]. Both mechanisms are diffusive. After the main event, which took place on the main fault, many surrounding small faults are activated, leading to a diffusion of the stress; a homogeneous diffusion of the stress could explain the monofractal distribution of the aftershock occurrences. Similar behaviors have been reported in [36].

**Figure Captions**

Fig. 1. Epicentral distribution of the seismicity of central Italy during the period 1983-2003.

Fig. 2. Interspike interval time series of the seismicity shown in Fig. 1.

Fig. 3. Fluctuation functions for q=-10 and q=10. The different slopes suggest the presence of multifractality in the series.

Fig. 4. Spectrum of the generalized Hurst exponents $h_q$ for q ranging between –10 and 10 (step of 0.5). The dependence of $h_q$ versus q is typical of multifractal sets.

Fig. 5. 3D plot of the time variation of the exponent $h_q \sim q$.

Fig. 6. Projection of the 3D plot of Fig. 5 on the plane $h_q$-t.

Fig. 7. Grey-scale map of the time variation of the generalized Hurst exponents $h_q$. The x axis is given by the event number. The vertical red line indicates the time occurrence of the largest shock of the series (M=5.8).

Fig. 8. Time variation of the range $\rho_q$ (a) and the standard deviation $\sigma_q$ (b) of the exponents $h_q$. There is also shown the time variation of the mean range $<\rho_{q,shuffled}>$ and the mean standard deviation $<\sigma_{q,shuffled}>$, obtained averaging the values over ten shuffles of the original series (see text for details).

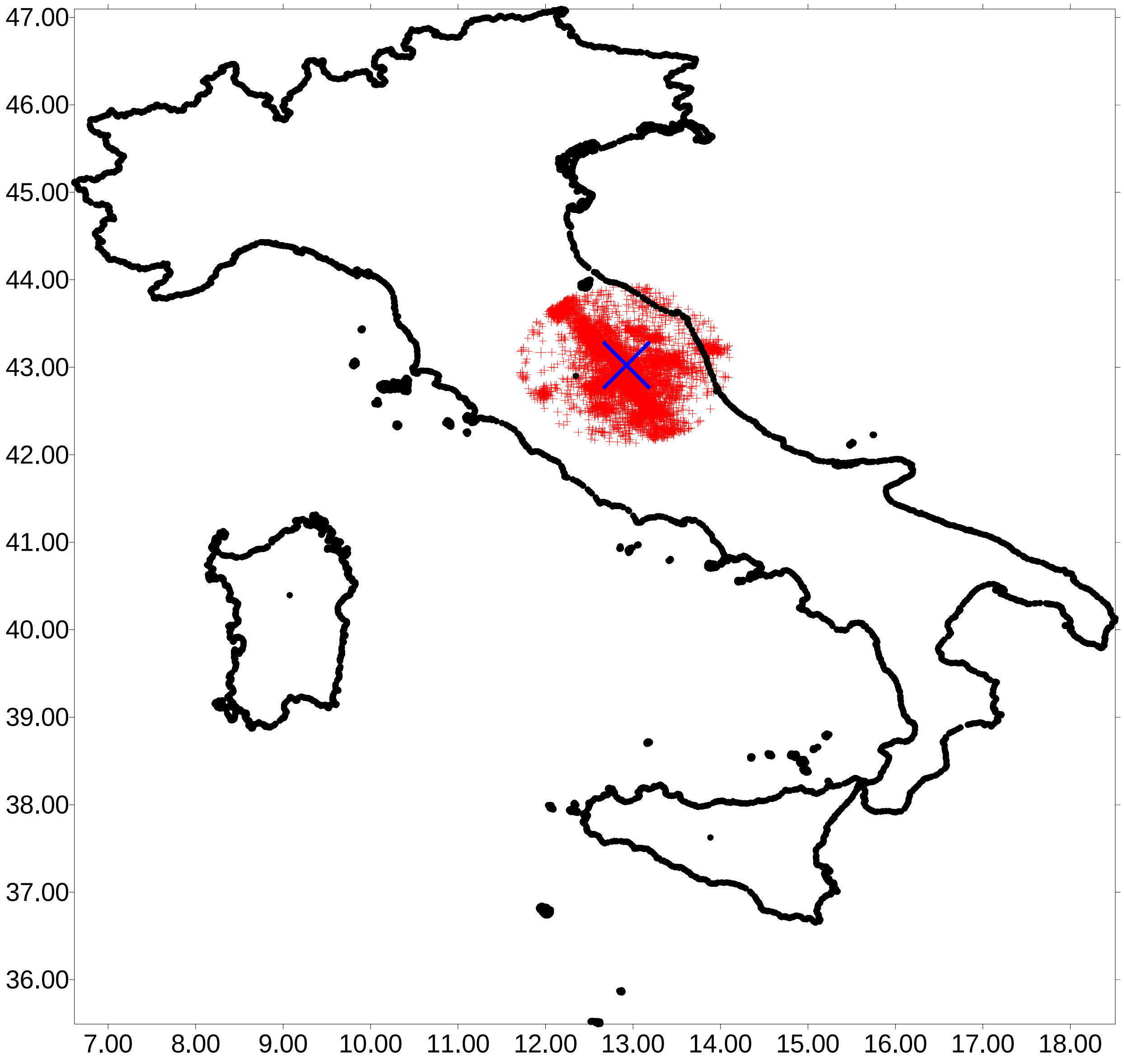

Fig. 1

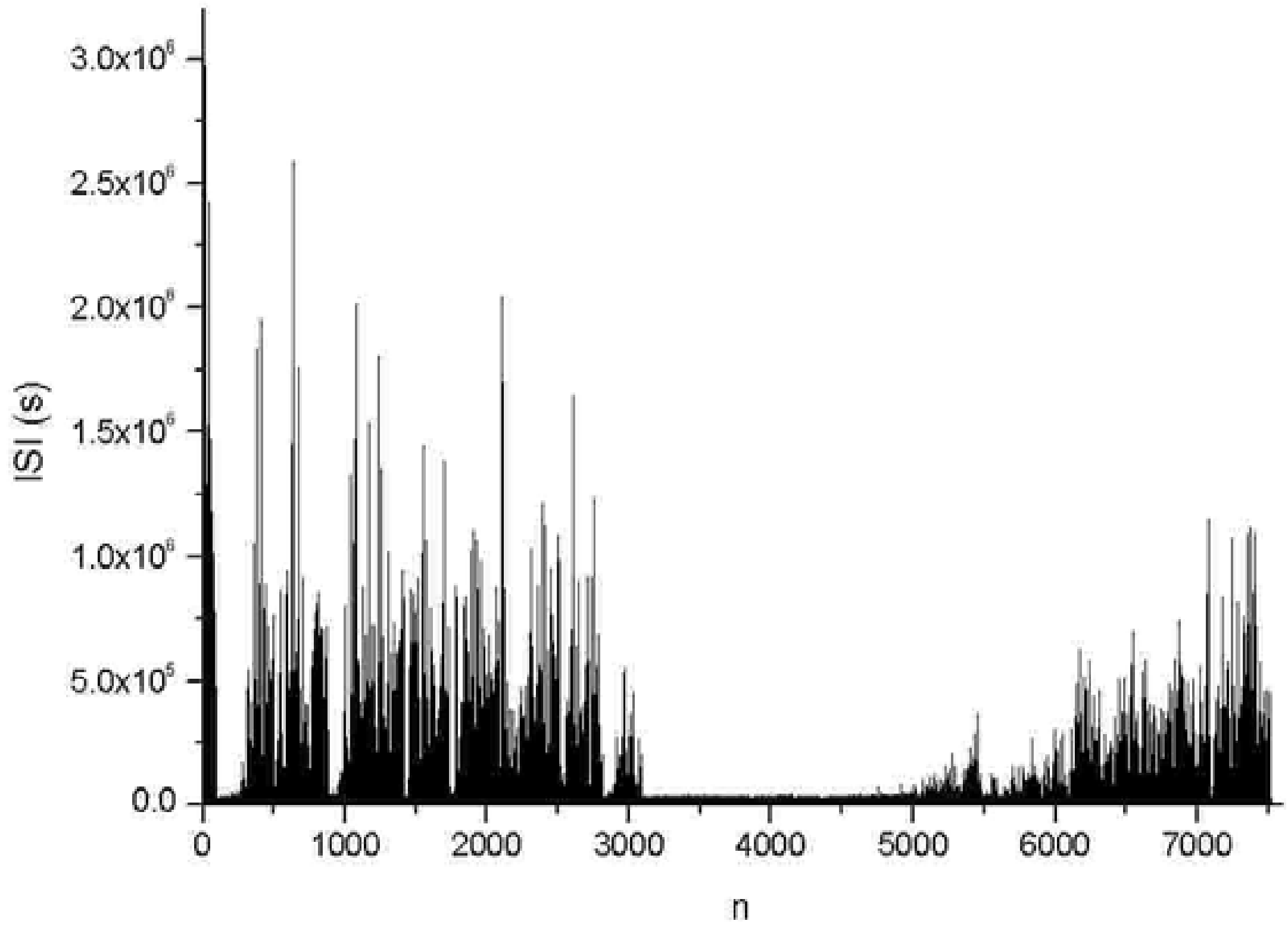

Fig. 2

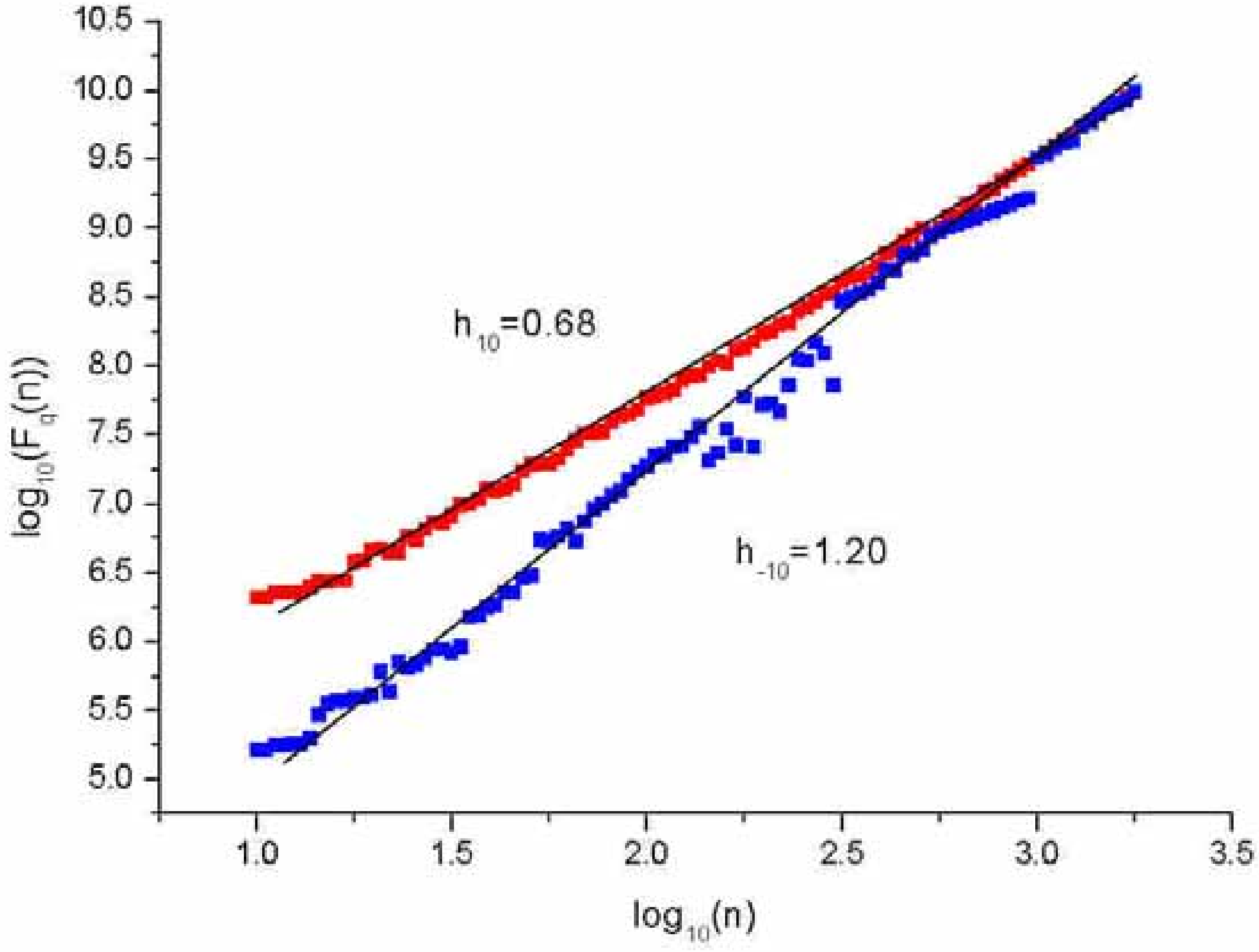

Fig. 3

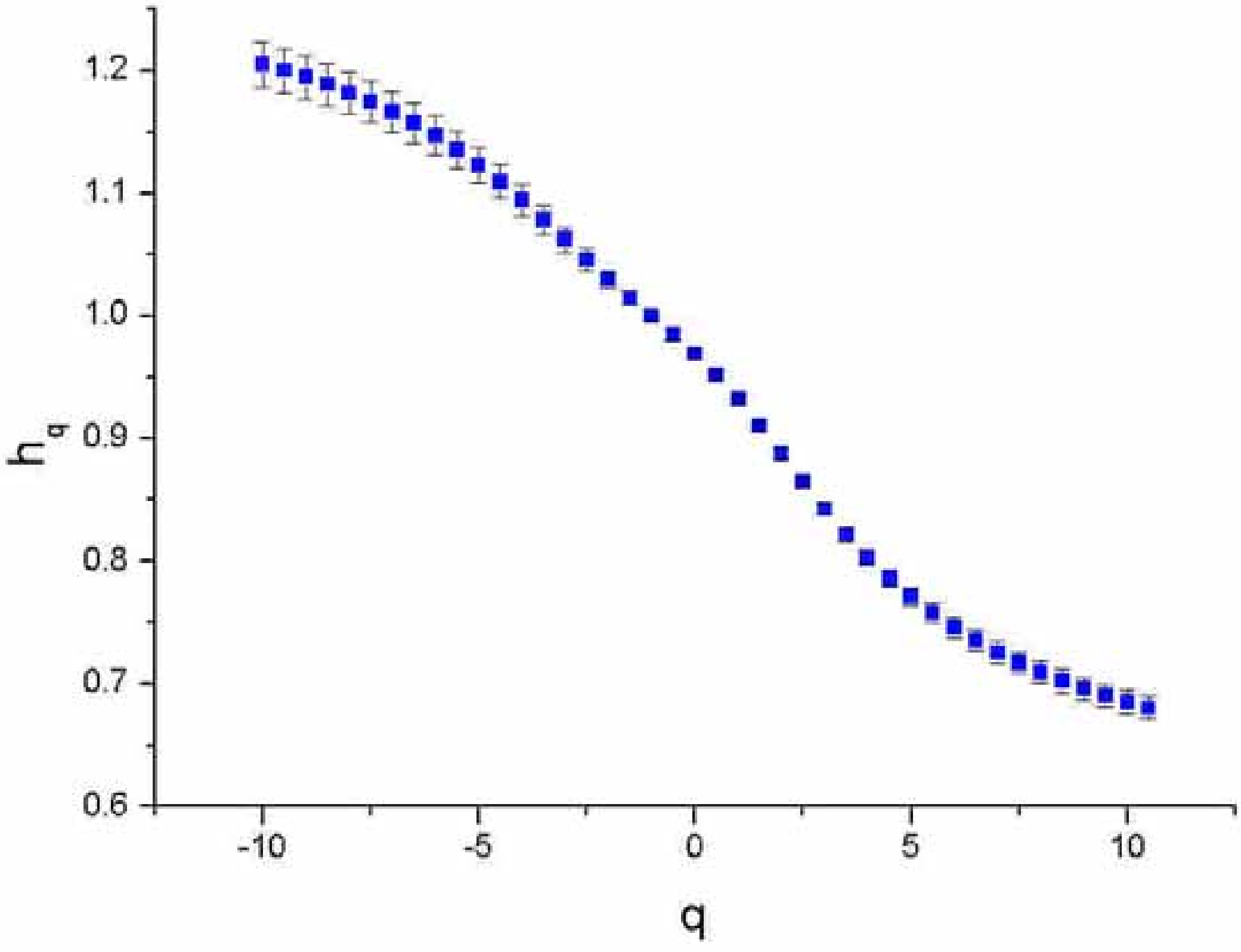

Fig. 4

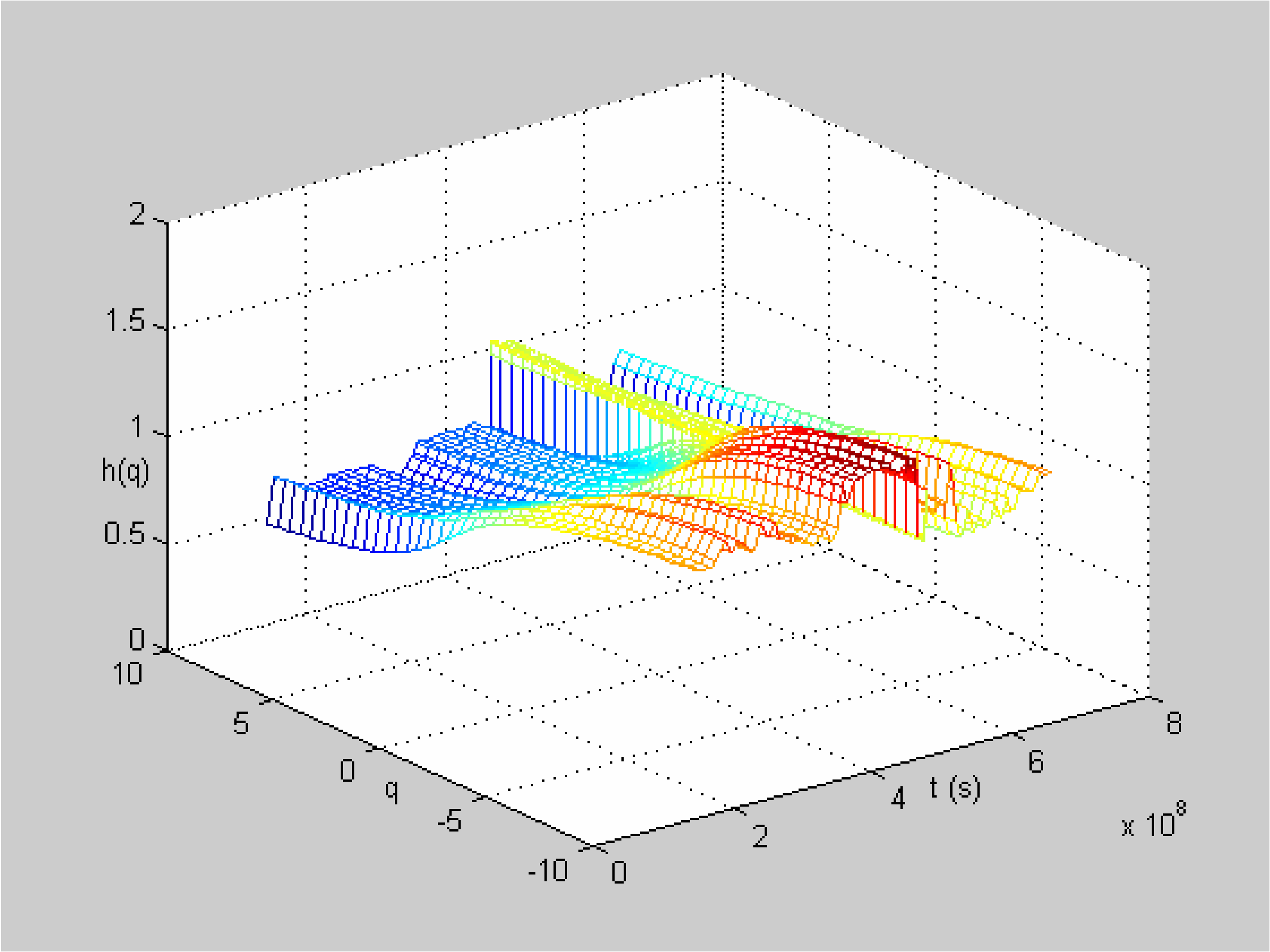

Fig. 5

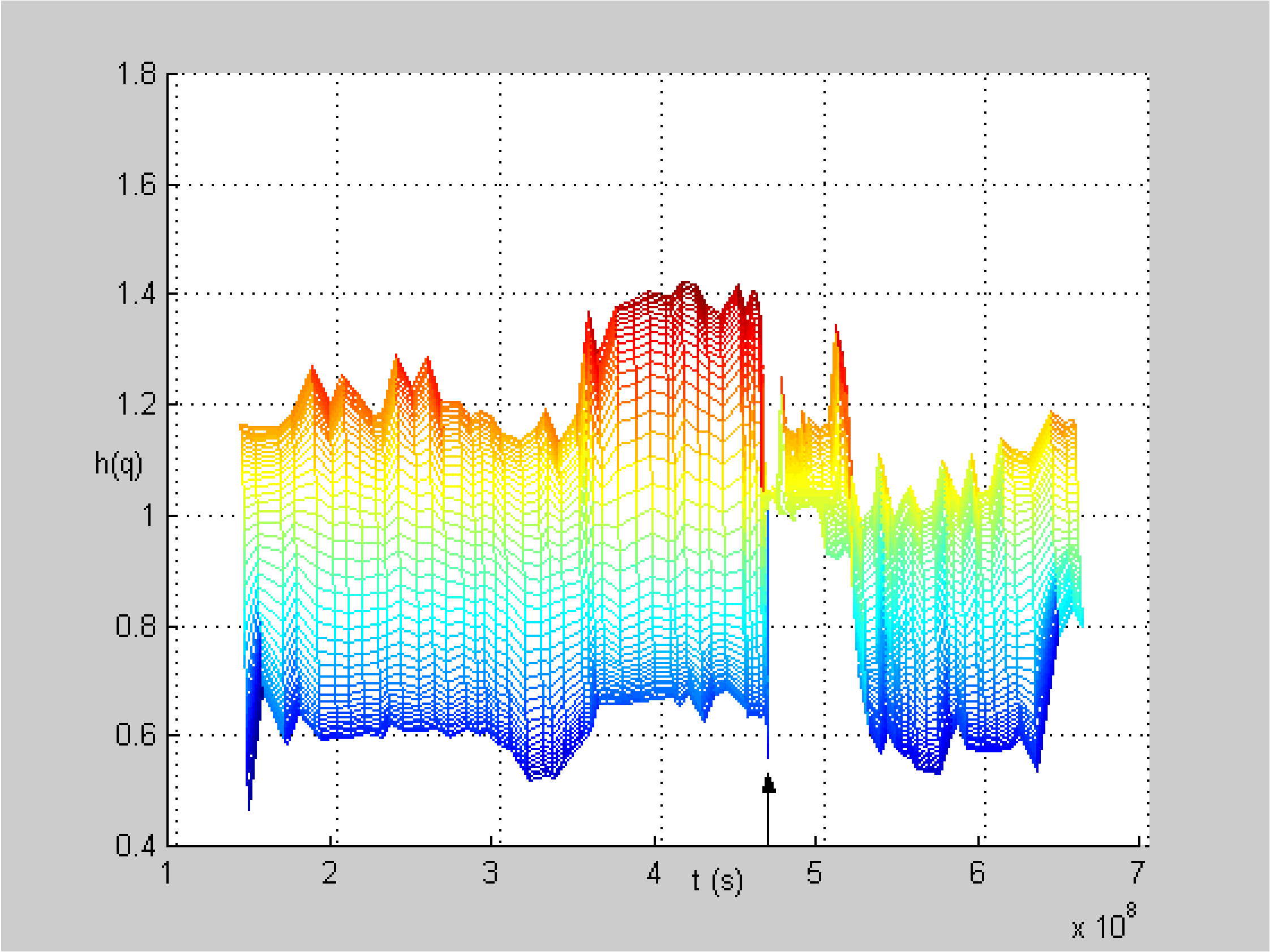

Fig. 6

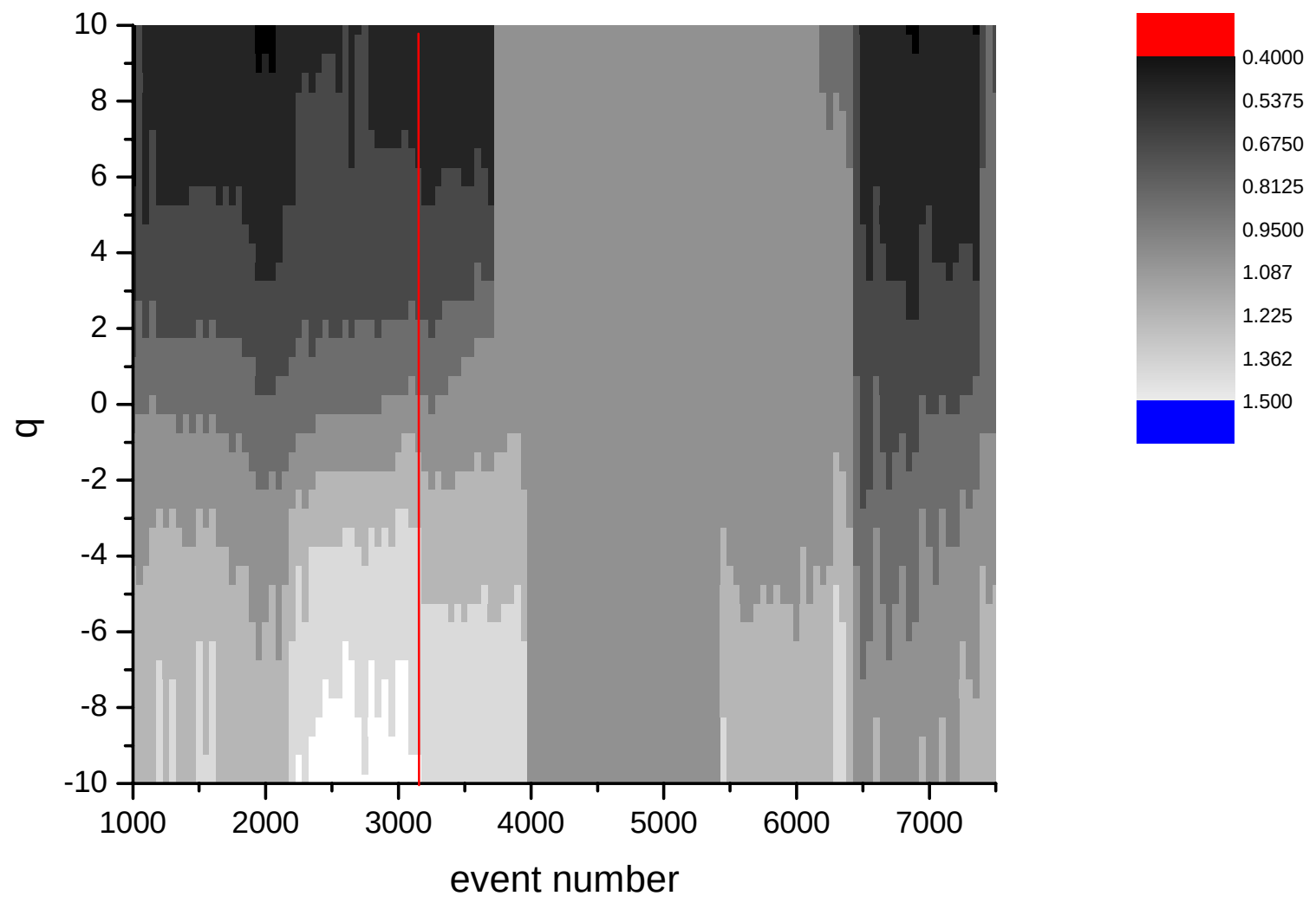

Fig. 7

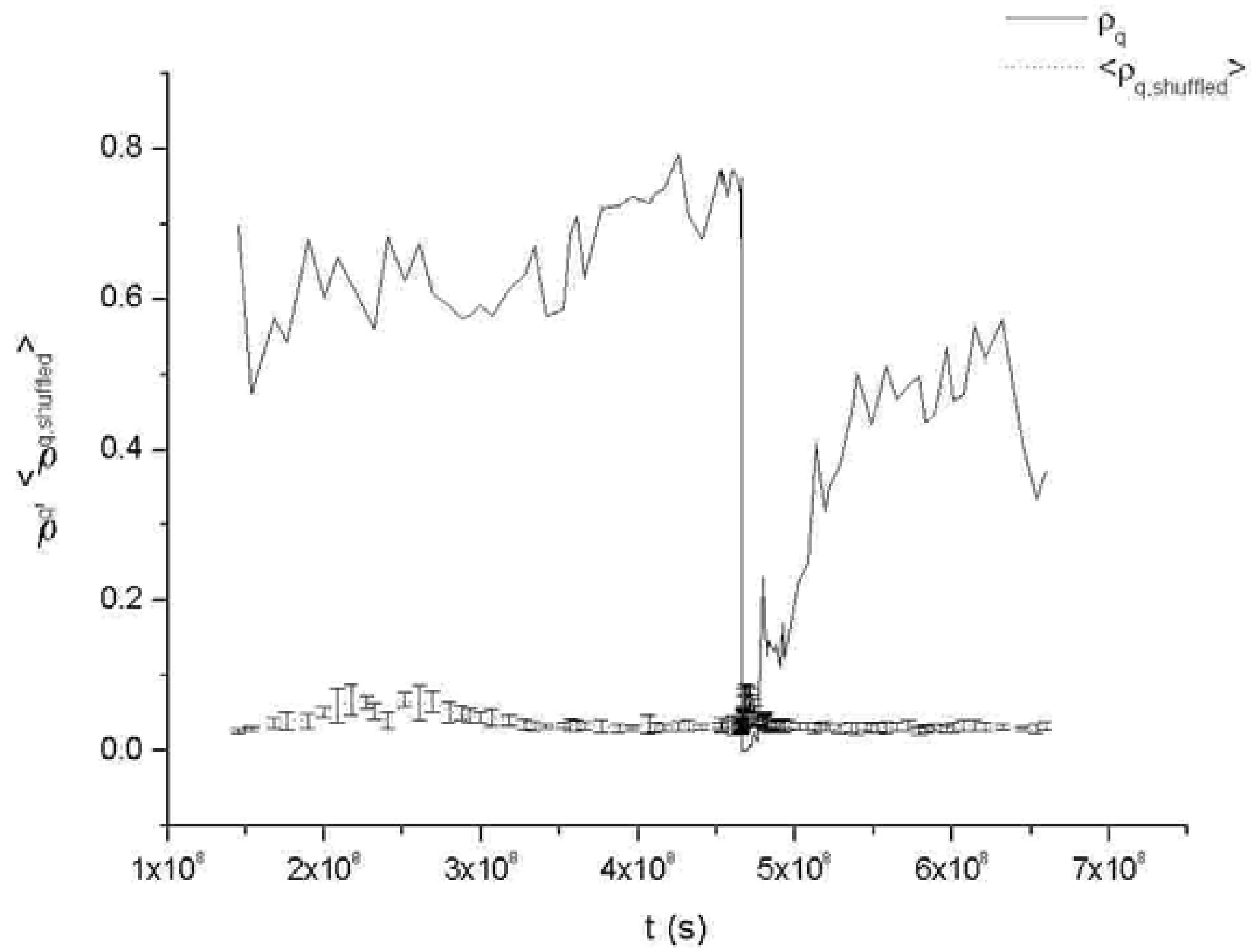

a)

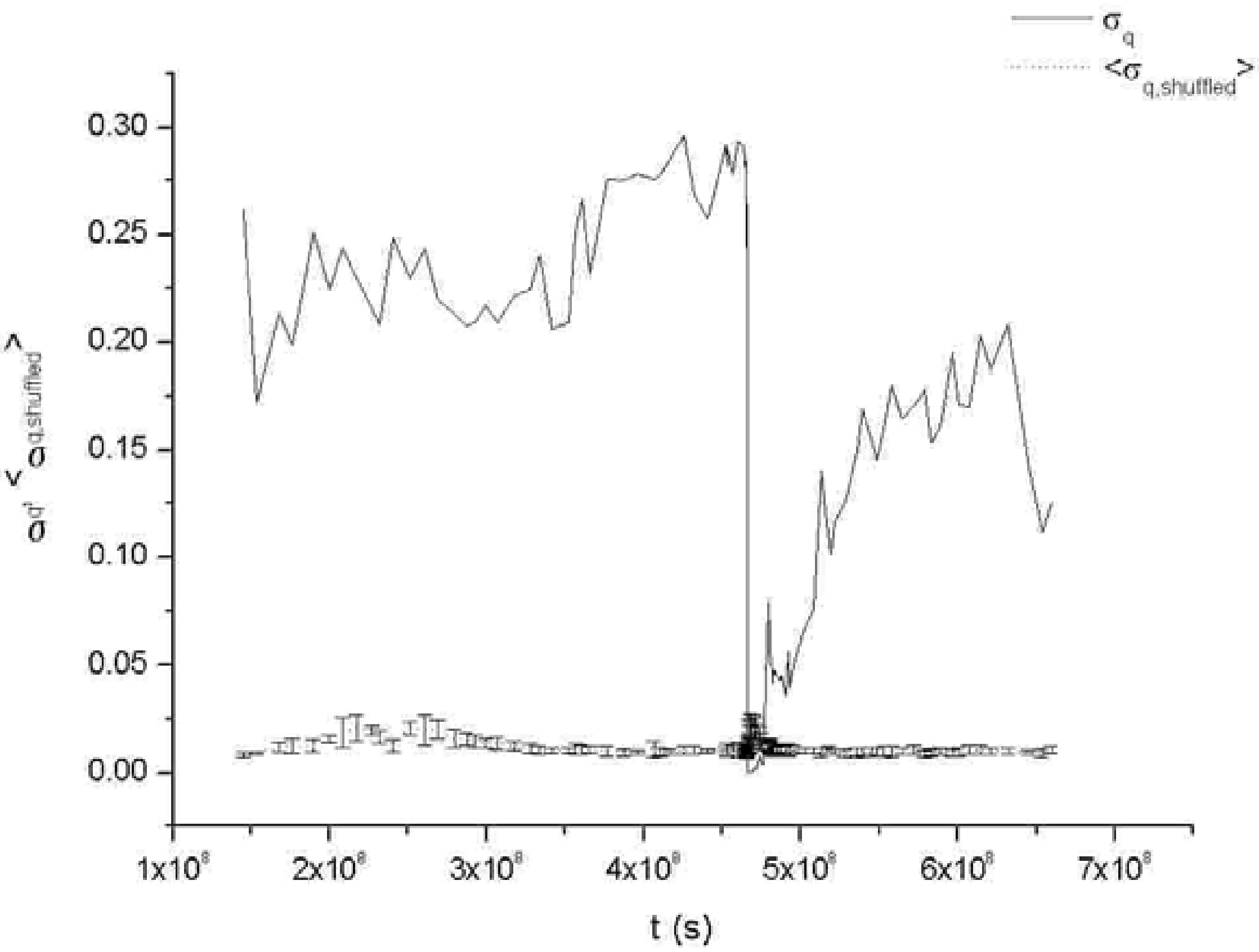

b)
Fig. 8